\documentclass[aps,showpacs,twocolumn,showkeys,amsmath,amssymb,floatfix,nofootinbib,preprintnumbers]
{revtex4}
\pdfoutput=1
\usepackage{bm}
\usepackage{epsfig}

\begin{document}

\title{The Pentaquark Candidates in the Dynamical Diquark Picture}

\author{Richard F. Lebed}
\email{richard.lebed@asu.edu}
\affiliation{Department of Physics, Arizona State University, Tempe,
Arizona 85287-1504, USA}

\date{July, 2015}

\begin{abstract}
  Starting with the dynamical picture of the exotic $c\bar
  c$-containing states $XY \! Z$ as the confinement-induced
  hadronization of a rapidly separating pair of a compact diquark and
  antidiquark, we describe the pentaquark candidates $P_c^+(4380)$ and
  $P_c^+(4450)$ in terms of a confined but rapidly separating
  color-antitriplet diquark $cu$ and color-triplet ``triquark'' $\bar
  c (ud)$.  This separation explains the relatively small $P_c^+$
  widths, despite these 5-quark systems lying far above both the
  $J/\psi \, p$ and $\Lambda_c \bar D^{(*)0}$ thresholds.  The $P_c^+$
  states are predicted to form isospin doublets with neutral partners
  $P_c^0$.
\end{abstract}


\maketitle


\section{Introduction} \label{sec:Intro}

The recent observation by LHCb~\cite{Aaij:2015tga} of prominent exotic
structures $P_c^+(4380)$ and $P_c^+(4450)$ in the $J/\psi \, p$
spectrum of $\Lambda_b \to J/\psi \, K^- p$ has rekindled hopes that
the long-sought pentaquark states have finally been observed.  The
reported properties are $m_1 = 4380 \pm 8 \pm 29$~MeV, $\Gamma_1 = 205
\pm 18 \pm 86$~MeV (at 9 standard deviations) and $m_2 = 4449.8 \pm
1.7 \pm 2.5$~MeV, $\Gamma_2 = 39 \pm 5 \pm 19$~MeV (at 12 standard
deviations), respectively, while the preferred $J^P$ assignments are
correlated, with the most likely combinations in decreasing order
being $( {\frac 3 2}^- \! , \, {\frac 5 2}^+ )$, $( {\frac 3 2}^+ \! ,
\, {\frac 5 2}^- )$, and $( {\frac 5 2}^+ \! , \, {\frac 3 2}^-)$.

Should at least one of the states be confirmed by another experiment,
it will join the famed tetraquarks---whose best-studied member [the
$J^{PC} = 1^{++}$ $X(3872)$] was discovered over a decade
ago~\cite{Choi:2003ue}---as a second class of exotic hadrons.  Since
the valence structure of $J/\psi \, p$ is $c\bar c uud$, the minimal
quark content of such states is that of a pentaquark.

We note several interesting phenomenological facts.  First, the two
charged {\em five}-quark hidden-charm states $P_c^+(4380)$ and
$P_c^+(4450)$ observed at LHCb are both lighter than the {\em
four}-quark state $Z^-(4475)$ [formerly $Z^-(4430)$] observed by the
same group~\cite{Aaij:2014jqa}.  Since pentaquarks carry baryon number
and therefore must have a baryon in their decay products, while
tetraquarks are bosons and therefore can decay entirely to
(generically lighter) mesons, less phase space is typically available
to $P_c^+$ decays.  In particular, $Z^-(4475)$ decays dominantly to
$\psi(2S) \, \pi^-$, even though plenty of phase space is available
for the $J/\psi \, \pi^-$ mode, while neither of the $P_c^+$ states is
kinematically allowed to decay to $\psi(2S) \, p$.

Second, both of the $P_c^+$ states lie well above the thresholds for
decay into $\Lambda_c^+ {\overline D}^{(*)0}$, which also has the
valence-quark structure $c\bar c uud$.  Since these two-body decays
are not forbidden by any obvious quantum number, the relatively small
$P_c^+$ widths suggest interesting internal structure sufficient to
suppress this immediate rearrangement, as well as that into $J/\psi \,
p$.  The $P_c^+(4450)$ lies only about 10~MeV below the $\Sigma_c^+
{\overline D}^{*0}$ threshold, which can be used to argue for a
molecular interpretation~\cite{Karliner:2015ina,Roca:2015dva}.
Alternate molecular assignments for the $P_c^+$ states are presented
in Refs.~\cite{Yang:2011wz,Chen:2015loa,Chen:2015moa,He:2015cea}, and
the origin of the $P_c^+(4450)$ as a threshold rescattering effect via
$\chi^{\vphantom{+}}_{c1} \, p$ is discussed in
Ref.~\cite{Guo:2015umn}, and through additional channels in
Ref.~\cite{Liu:2015fea}.

Third, all of the preferred fits from LHCb demand that the two $P_c^+$
states carry opposite parities.  A system of four quarks and one
antiquark in a relative $S$ wave has negative parity, while positive
parity requires the introduction of at least one unit of relative
orbital angular momentum ($P$ wave or higher).  And yet, the two
$P_c^+$ states are separated by only $m_2 - m_1 = 70$~MeV\@.  One may
argue that the figure of merit, as would be the case for Regge
trajectories, should be $m_2^2 - m_1^2 = (790 \ {\rm MeV})^2$, which
is a much more natural hadronic scale; however, attempting to discern
a trajectory when only two points are available seems absurdly
premature.  Still, if the $J^{PC} = 1^{--}$ state $Y(4008)$ seen by
Belle with a mass $3891 \pm 42$~MeV~\cite{Liu:2013dau} is confirmed,
then small mass splittings between hidden-charm exotics of opposite
parities [{\it i.e.}, only 20~MeV from the $X(3872)$]---and hence
relative orbital excitations---will not appear unusual.

In this short paper, we propose a structure for the $P_c^+$ states
based upon a mechanism recently proposed for understanding the
tetraquark states~\cite{Brodsky:2014xia}, wherein a diquark $\delta$
and an antidiquark $\bar \delta$ form in the attractive color-$\bar
{\bf 3}$ and color-${\bf 3}$ representations, respectively.  This
configuration is prevented from instantly reorganizing into two $\bar
q q$ pairs because the diquarks are forced to separate rapidly due to
the large energy release in the production mechanism ({\it e.g.}, a
$B$-meson decay via $b \to c \bar c s$).  However, since the diquarks
are colored, they are confined and hence cannot separate indefinitely;
kinetic energy is converted into the potential energy of a color flux
tube connecting them.  Hadronization finally occurs through the
overlap of the long-distance tails of hadronic wave functions that
stretch from the quarks in $\delta$ to the antiquarks in $\bar
\delta$.  This picture was used to explain, for example, the
preference for $Z^-(4475) \to \psi (2S) \, \pi^-$ over $Z^-(4475) \to
J/\psi \, \pi^-$, simply due to the final spatial separation of the
$c$ in $\delta$ and the $\bar c$ in $\bar \delta$ allowing a much
greater wave function overlap with the larger $\psi(2S)$ than with the
more compact $J/\psi$.

We argued implicitly in~\cite{Brodsky:2014xia}, and explicitly in a
subsequent paper~\cite{Blitz:2015nra}, that the diquarks formed from
one heavy and one light quark are somewhat smaller than diquarks
formed from two light quarks, and that such heavy-light $(Qq)$
diquarks are not expected to be much larger than the meson $(Q\bar q)$
formed from the same flavors.  In yet another subsequent
paper~\cite{Brodsky:2015wza}, we proposed that the preferential
coupling of ${\bf 3} \otimes {\bf 3} \to \bar {\bf 3}$ (and its
conjugate) does not end with diquarks, but can continue sequentially
to more complex structures like pentaquarks and even octoquarks.
These ideas will be used to develop the pentaquark picture described
in detail below.  As in Ref.~\cite{Brodsky:2014xia}, we use the term
``picture'' because many distinct models could be constructed that
support this dynamics, and want to emphasize that this discussion is
not limited to any particular choice of potential or wave functions,
for example.

The concept that pentaquarks might be formed from two compact colored
constituents rather than molecules of mesons was first expressed in
Ref.~\cite{Karliner:2003dt}, which sought to describe the $\Theta^+
(1535)$ pentaquark candidate $u \bar s udd$ as a molecule of a $(ud)$
diquark in a color-$\bar {\bf 3}$ and a $(\bar s u d)$ ``triquark'' (a
term coined in that paper) formed from a $(ud)$ diquark in a
color-${\bf 6}$ coupled to the $\bar s$ quark into an overall
color-${\bf 3}$.  The separation of the two components of the molecule
is stabilized by the centrifugal barrier introduced by relative
orbital angular momentum $\ell = 1$, which as discussed above was
necessary should the $\Theta^+$ have been found to carry positive
parity.  In comparison, the famous $\Theta^+$ pentaquark model of
Ref.~\cite{Jaffe:2003sg} proposed a structure of two light $(ud)$
diquarks and one exceptional ($\bar s$) quark.  Both
Refs.~\cite{Karliner:2003dt} and \cite{Jaffe:2003sg}, make use of
diquarks consisting of light quarks in the ``good'' (spin-0)
combinations, so named because they are believed to be more tightly
bound than the ``bad'' (spin-1) combination through hyperfine
couplings (although in Ref.~\cite{Karliner:2003dt} the diquark inside
the triquark has spin 1).  In the case of heavy-light diquarks, the
hyperfine couplings are proportional to $1/m_Q$, and therefore the
mass difference between ``good'' and ``bad'' is greatly reduced.
Alternate compositions through colored components have been very
recently discussed in Refs.~\cite{Mironov:2015ica,Maiani:2015vwa}.

This paper is organized as follows: In Sec.~\ref{sec:Diquark} we
discuss the diquark picture used to describe the new exotic states.
Section~\ref{sec:PentaPic} presents the diquark picture as relevant to
the production of the $P_c^+$ states.  In Sec.~\ref{sec:Phenom} we
present the basic phenomenology for the $P_c^+$ states provided by
this picture, and in Sec.~\ref{sec:Concl} we summarize.

\section{Generalizing the Diquark Picture} \label{sec:Diquark}

The color algebra of QCD provides more than one way to obtain
attractive channels between quarks.  The combination ${\bf 3} \otimes
\bar {\bf 3} \to {\bf 1}$, which explains the strong binding of $q\bar
q$ pairs in conventional mesons, is of course extremely well known.
However, one other binary combination is strongly binding, the channel
${\bf 3} \otimes {\bf 3} \to \bar {\bf 3}$.  To quantify the effect,
note that the coupling of two colored objects in the irreducible
representations $R_1$ and $R_2$ is computed by the same techniques as
one computes the spin coupling between objects carrying spins $S_1$
and $S_2$ combining to total spin $S = |{\bf S}_1 + {\bf S}_2|$, via
the trick
\begin{equation} \label{eq:spinspin}
{\bf S}_1 \cdot {\bf S}_2 = \frac 1 2 \left[ \left( {\bf S}_1 + {\bf
S}_2 \right)^2 - {\bf S}_1^2 - {\bf S}_2^2 \right] \, .
\end{equation}
One generalizes to an arbitrary Lie algebra by computing the
combination of quadratic Casimirs
\begin{equation} \label{eq:Casimirs}
g_{1 \times 2} \equiv C_2 (R) - C_2 (R_1) - C_2 (R_2) \, .
\end{equation}
Considering all binary combinations ${\bf 3} \otimes {\bf 3} = \bar
{\bf 3} \oplus {\bf 6}$ and ${\bf 3} \otimes \bar {\bf 3} = {\bf 1}
\oplus {\bf 8}$, one computes the relative strengths
\begin{equation} \label{eq:colors} g_{1 \times 2} = \frac 1 3 ( -8,
  -4, +2, +1) \ {\rm for} \ R = ( {\bf 1}, \bar {\bf 3}, {\bf 6} ,
  {\bf 8} ) \, .
\end{equation}
The diquark attractive $\bar {\bf 3}$ channel $\delta$ is therefore
fully half as strong as that of the singlet $q \bar q$ channel.  In a
multiquark system in which two quarks or two antiquarks happen to lie
in closer proximity than to one of their antiparticles, the diquark
$\delta$ (antidiquark $\bar \delta$) attraction is naturally expected
to dominate.  Unless stronger color forces intervene ({\it e.g.}, the
production of nearby $\bar q$'s, which would create available
color-singlet combinations), the $\delta$ and $\bar \delta$
combinations can be expected to form quasi-bound, but colored and
therefore confined, states.

In Ref.~\cite{Brodsky:2014xia}, the diquarks containing a charm and a
light quark were crudely estimated to have a comparable size to a $D$
meson, roughly $\left< r \right> \alt 0.5$~fm.  The subsequent paper
Ref.~\cite{Blitz:2015nra} argued that diquarks formed with a heavy
quark should be somewhat smaller than those formed from two light
quarks: A heavier quark is more localized in space, while each lighter
quark has a more diffuse wave function.  A key phenomenological
question in identifying whether diquarks have formed is whether or not
any antiquarks appear within this radius.

This hypothesis was used in Ref.~\cite{Brodsky:2015wza} to suggest a
means by which multiquark exotics could be produced, particularly at
threshold, where the limited phase space allows the soft heavy quark
pairs such as $c\bar c$ to coalesce with light valence quarks moving
at similar rapidities.  Such multiquark states can be formed through a
sequence of two-body bound-state clusters of color-$\bar {\bf 3}$
diquark and color-${\bf 3}$ antidiquark states.  In the absence of
easy opportunities for the formation of color singlets, sequential
diquark formation provides the strongest channels for binding.  For
example, anti-de~Sitter/QCD models on the light
front~\cite{Brodsky:2014yha} have a universal confining potential that
confirms the importance of diquarks in hadron spectroscopy.  While the
examples given in Ref.~\cite{Brodsky:2015wza} describe literal
clusters of diquarks such as a charmed, charge $Q = 4$, baryon-number
$B = 2$ state $[uu]_{\bar 3C} [cu]_{\bar 3C} [uu]_{\bar 3C}$, another
route of sequential color-triplet (antitriplet) formation is
available, which is the hypothesis of this paper: A pre-existing
diquark $\delta^\prime$ that subsequently encounters an antiquark
$\bar Q$ forms a bound {\it antitriquark\/} $\bar \theta \equiv ( \bar
Q \delta^\prime)$ via the attractive color coupling $\bar {\bf 3}
\otimes \bar {\bf 3} \to {\bf 3}$.  This mechanism, as discussed in
the next section, provides a completely analogous production channel
for pentaquark states to that described for tetraquark states in
Ref.~\cite{Brodsky:2014xia}.

To say that two quarks or antiquarks encountering one another combine
only into the most attractive channel is of course a great
simplification.  First, the color coupling factors apply without
reservations only when fundamental QCD interactions dominate the
interaction.  Longer-distance effects dress the interactions and can
obfuscate this simple result.  In reality, one expects a type of
thermodynamic ensemble of states in various color combinations, in
which the levels at the lowest energies are driven by diquark binding.
The existence of such an ensemble assumes that the formation of
overall color singlets is precluded due to the presence of a potential
energy barrier, such as from a large spatial separation between the
quarks needed to form the singlets.  In that case, the eventual
hadronization can be considered as a tunneling process.  Second, the
Pauli exclusion principle must be taken into account if the purported
diquarks contain quarks of identical flavor, since then the flavor
wave function is automatically symmetric, and therefore the color-spin
wave function must be antisymmetric.  In the specific example
discussed below, this constraint is not an issue, but it must be kept
in mind for other cases.

\section{Pentaquark Production Mechanism} \label{sec:PentaPic}

\begin{figure}
\begin{center}
\includegraphics[width=3.6in]{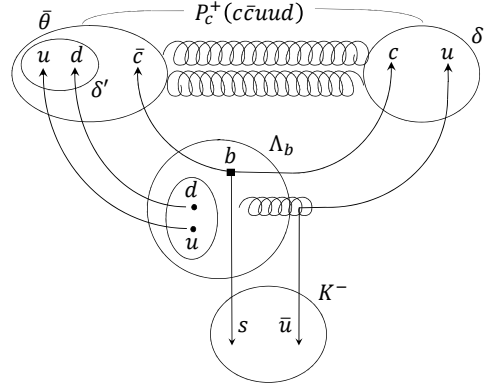}
\caption{Illustration of the production of a spatially extended
  diquark-antitriquark state $\delta \bar \theta$ attracted by
  long-range color forces (indicated by gluon lines) via a color flux
  tube.  Here, the mechanism is illustrated for $\Lambda_b \to P_c^+
  K^-$, where the black square indicates the $b$-quark weak decay.}
\label{Fig:Pc}
\end{center}
\end{figure}

We propose that the states $P_c^+$ observed at LHCb are pentaquarks
consisting of a confined but rapidly separating pair of a color-$\bar
{\bf 3}$ diquark $\delta = (cu)$ and a color-{\bf 3} antitriquark
$\bar \theta = \bar c (ud)$, in which the $(ud)$ subsystem of $\bar
\theta$ is a color-$\bar {\bf 3}$ diquark $\delta^\prime$, as depicted
in Fig.~\ref{Fig:Pc}.

This picture is completely analogous to (the charge conjugate of) that
for the $Z^+(4475)$ presented in Ref.~\cite{Brodsky:2014xia}, except
that the diquark $\delta^\prime = (ud)$ in Fig.~\ref{Fig:Pc} is
replaced by the single quark $\bar d$.  The parent hadron for the
$P_c^+$ is the $\Lambda_b$ baryon, while the parent hadron for the
$Z_c^+$ is the $\bar B^0$.  In either case, the composite state is not
a molecule in the traditional sense of the word, because it lasts only
as long as the $\delta$-$\bar \theta$ pair ($\delta$-$\bar \delta$ in
Ref.~\cite{Brodsky:2014xia}) possess positive kinetic energy to
continue separating.  The colored $\delta$ and $\bar \theta$
constituents create a color flux tube between them, losing their
energy to the color field and slowing down.  For this picture to be
physically meaningful, the $\delta$-$\bar \theta$ pair must be
sufficiently compact that their wave function overlap becomes
insignificant.  As in the standard Wentzel-Kramers-Brillouin
approximation, the probability of a transition---in this case,
hadronization---increases as the components approach the classical
turning point.

Let us emphasize the differences between this and other pictures for
the pentaquark.  First, it is clearly different from the hadronic
molecule picture, in which the constituent baryon and meson are color
singlets treated as forming a static molecule and are held together by
weak color van der Waals forces.  It is also different from the
diquark-triquark model of Ref.~\cite{Karliner:2003dt}, not only
because the diquarks are always assumed here to form color triplets,
but also because the diquark-triquark state in that case is again a
static molecule, stabilized by a centrifugal barrier; in our picture,
such states are expected to last only as long as the components
continue to separate, and also would exist in $S$ waves as well as
higher partial waves.  The recent work of Ref.~\cite{Maiani:2015vwa}
describes the $P_c^+$ states as being formed of diquarks, via the
composition $\bar c_{\bar {\bf 3}} (cq)_{\bar {\bf 3}} (q^\prime
q^{\prime\prime})_{\bar {\bf 3}}$; while the color structure is the
same as ours, we emphasize the importance of the $\bar c$ belonging to
a compact component of the overall state.

The new ingredient for the $P_c^+$ as compared to the $Z^+(4475)$ is
the use of the intrinsic diquark $\delta^\prime = (ud)$ originating in
the $\Lambda_b$.  Note from Fig.~\ref{Fig:Pc} that $\delta^\prime$
acts as a spectator in the $P_c^+$ production process.  $\Lambda_Q$
baryons have always had a special place in the history of diquark
models, because these baryons by definition are isosinglets, and since
the heavy quarks $Q = s, c, b$ are also isosinglets, the remaining
quark pair $(ud)$ also forms an isosinglet, with a wave function that
is antisymmetric under flavor exchange.  Since $Q$ is a color-{\bf 3},
$(ud)$ is a color $\bar {\bf 3}$, again an antisymmetric combination.
The Pauli principle therefore demands an antisymmetric spin wave
function for $(ud)$.  Since both the ground-state $\Lambda_Q$ baryons
and the heavy quarks $Q$ have $J^P = {\frac 1 2}^+$, the $(ud)$ is
therefore expected to live in the antisymmetric spin-0 combination.
The $(ud)$ pair in $\Lambda_Q$ baryons is frequently termed a diquark,
and indeed it has exactly the color structure we want; it differs from
the ones we have previously discussed only by consisting solely of
light quarks, and is in the ``good'' diquark combination only.

As mentioned above, one may expect the $(ud)$ diquark to be slightly
larger than the heavy-light diquarks, but even so, its binding to the
heavy quark $b$ in $\Lambda_b$ and $c$ in $\bar \theta$ restricts the
full spatial extent of the wave function.  For example, using
heavy-quark symmetry and a variational approach,
Ref.~\cite{Albertus:2003sx} calculate root-mean square matter radii
for $\Lambda_b$ and $\Lambda_c$ to be no more than 0.22~fm and
0.31~fm, respectively.  Treating the $\bar \theta$ antitriquark as a
``would-be'' $\Lambda_c$ baryon ({\it i.e.}, differing by $\bar c
\leftrightarrow c$ but otherwise bound by essentially the same
nonperturbative physics), one expects $\bar \theta$ to be not much
larger than $\Lambda_c$.  In principle, the $u$ quark created from a
gluon can mix with the one in the $\Lambda_b$ diquark.  However,
inasmuch as this initial $(ud)$ diquark is expected to be fairly
tightly bound, one expects it to propagate as an undisturbed spectator
quasiparticle through the process; otherwise, the most likely outcome
would be a dissociation of the diquark, leading to a different
intermediate state than described in Fig.~\ref{Fig:Pc}.

In the case of the observed decay $P_c^+ \to J/\psi \, p$, the decay
rate is suppressed due to the final separation of the $c$ quark (in
$\delta$) and $\bar c$ quark (in $\bar \theta$) compared to the
typical size of the $J/\psi$ wave function, and to a lesser extent due
to the separation of the $(ud)$ diquark $\delta^\prime$ in $\bar
\theta$ from the $u$ quark in $\delta$ compared to the typical size
$\left< r_p \right> \simeq 0.88$~fm of the proton wave function.

\section{Phenomenology of the $P_c^+$ States} \label{sec:Phenom}

The first interesting point one notes from Fig.~\ref{Fig:Pc} is the
peripheral role played in the process by the $u \bar u$ pair created
by gluodynamics.  Certainly, creation instead of a $d \bar d$ pair
would give a nearly identical scenario.  One therefore predicts
isodoublet partners $P_c^0$ to be produced via $\Lambda_b \to P_c^0
{\bar K}^0 \to J/\psi \, n {\bar K}^0$ at masses just a few MeV higher
(from $u \to d$) than those for the $P_c^+$ states.  That the $P_c$
states should form isospin doublets is of course guaranteed by $P_c^+
\to J/\psi \, p$ being a strong decay and hence conserving isospin,
while $I_{J/\psi} = 0$ and $I_p = 1/2$; nevertheless, it is useful to
see the charge symmetry process explicitly in the context of a
particular decay mechanism.

We now obtain a crude estimate of the separation of the diquark and
antitriquark using the same technique as in
Ref.~\cite{Brodsky:2014xia}: Since the two components transform as a
color-$({\bf 3}, \bar {\bf 3})$ pair, one may describe them using the
well-known linear-plus-Coulomb ``Cornell'' nonrelativistic
potential~\cite{Eichten:1978tg}.  In the most thorough recent
analysis~\cite{Barnes:2005pb}, the central part of the potential for
$c\bar c$ systems is given by
\begin{equation} \label{eq:Cornell}
V(r) = -\frac 4 3 \frac{\alpha_s}{r} + b r + \frac{32\pi\alpha_s}
{9m_c^2} \left( \frac{\sigma}{\sqrt{\pi}} \right)^3 \! \!
e^{-\sigma^2 r^2} {\bf S}_c \! \cdot {\bf S}_{\bar c} \, ,
\end{equation}
with $\alpha_s = 0.5461$, $b = 0.1425$~GeV$^2$, $m_c = 1.4797$~GeV,
and $\sigma = 1.0946$~GeV.  The $-4/3$ color factor is the same one as
in Eq.~(\ref{eq:colors}), and $c (\bar c)$ now refer to the
components containing these quarks, in our case $\delta$ and $\bar
\theta$, respectively.

The calculation of Ref.~\cite{Brodsky:2014xia} further exploited the
fact that at least one of the components $\delta$ and $\bar \delta$ is
in a state of zero spin for each state of interest [$X(3872)$ and
$Z^-(4475)$], so that ${\bf S}_c \cdot {\bf S}_{\bar c} = 0$, and that
both are expected to have all quarks in a relative $S$ wave, so that
noncentral contributions to $V(r)$ are not needed.  In the case of the
$P_c^+$ states, we have seen that at least one of them must have an
orbital excitation, so this calculation strictly applies only to the
$S$-wave state.  Furthermore, the antitriquark $\bar \theta$
necessarily carries half-integer spin; nevertheless, the ${\bf S}_c
\cdot {\bf S}_{\bar c}$ term has little effect on $V(r)$ except at the
smallest values of $r$, so we neglect it here.  Lastly, we use the QCD
sum-rule based estimate~\cite{Kleiv:2013dta} $m_\delta = 1.860$~MeV
(note its nearness to the $D^0$ mass 1.865~GeV) and the antitriquark
mass estimate $m_{\bar \theta} = m_{\Lambda_c} = 2.286$~GeV.  Using
these assumptions, the diquark-antitriquark separations $R$ obtained
from Eq.~(\ref{eq:Cornell}) are
\begin{eqnarray}
R = 0.64 \ {\rm fm} & {\rm for} & P_c^+(4380) \, , \nonumber \\
R = 0.70 \ {\rm fm} & {\rm for} & P_c^+(4450) \, . 
\label{eq:separation}
\end{eqnarray}
These distances are not especially large for light hadronic systems.
However, inasmuch as diquarks, and especially triquarks, containing
heavy quarks may be rather smaller as discussed above, these
components may be considered as well separated for the purpose of
computing quantum-mechanical wave function overlaps.  This separation,
particularly of the $c$ and $\bar c$ quark, explains the suppressed
decay rate to $J/\psi$, since the potential Eq.~(\ref{eq:Cornell})
gives $\left< r_{J/\psi} \right> = 0.39$~fm.  The similarly small size
for the $\Lambda_c$ also predicts slow transitions to $\Lambda_c^+
{\overline D}^{(*)0}$, and hence overall widths that are suppressed
compared to naive expectations.

Finally, let us consider the quantum numbers of the allowed states in
this picture.  Orbital excitations can occur not only along the flux
tube, but within the diquark and antitriquark as well.  Nevertheless,
let us for simplicity ignore the latter.  Inasmuch as the diquark
$\delta^\prime$ in $\bar \theta$ inherited from the $\Lambda_b$ has
spin zero, the set of allowed quantum numbers is even simpler, since
then the spin of $\bar \theta$ is $\frac 1 2$.  For $S$ waves, one has
the $J^P$ possibilities
\begin{equation} \label{eq:Swaves}
{\frac 1 2}^- \otimes 0^+ \otimes \left\{ \begin{array}{c} 0 \\ 1
\end{array} \right\}^+ = \left\{ \begin{array}{c} \frac 1 2 \\
\frac 1 2 \oplus \frac 3 2  \end{array} \right\}^- \, ,
\end{equation}
and for $P$ waves, one has
\begin{equation} \label{eq:Pwaves}
{\frac 1 2}^- \otimes 1^- \otimes \left\{ \begin{array}{c} 0 \\ 1
\end{array} \right\}^+ = \left\{ \begin{array}{c} \frac 1 2 \oplus
\frac 3 2 \\ \frac 1 2 \oplus \frac 3 2 \oplus \frac 5 2 \end{array}
\right\}^+ \, ,
\end{equation}
where the three numbers on the left-hand side are the spin of the
$\bar c (ud)$ antitriquark, the orbital excitation, and the spin of
the $(cu)$ diquark.

In this simplified picture, we find only one state with $J = \frac 5
2$, namely, the $P$-wave ${\frac 5 2}^+$.  It is natural to identify
this state with the higher-mass $P_c^+(4450)$, since it has a narrower
width that can be explained by the near-threshold phase-space
suppression of $P$ waves.  Then the broader $P_c^+(4380)$ must have
$J^P = {\frac 3 2}^-$ and lie in an $S$ wave.

Note that both of these states have the $(cu)$ diquark in a spin-1
configuration.  It is an interesting phenomenological fact of the
tetraquark sector that no $J^P = 0^+$ state has yet been confirmed; in
the context of the diquark-antidiquark picture, the simplest such
states would have both diquarks in $S=0$ combinations, with $L=0$ in
the color flux tube as well.  It is plausible that such states are
much broader due to the absence of any angular momentum barriers
impeding rapid decays.  Likewise, the lower-spin states in
Eqs.~(\ref{eq:Swaves}) and (\ref{eq:Pwaves}) might be more difficult
to discern experimentally.  In any case, the discovery of two new
states and the possibility of numerous others left to find will
certainly spur on further experimental examination.

\section{Conclusions} \label{sec:Concl}

We have seen that the recently observed charmoniumlike pentaquark
candidates $P_c^+$ may have a common dynamical origin with the
charmoniumlike tetraquark states.  Both are proposed to occur as
systems of rapidly separating color-{\bf 3} and -$\bar {\bf 3}$
component pairs, and in particular in the $P_c^+$ pentaquarks through
the sequential preferential formation of color-triplet combinations
$[\bar c(ud)_{\bar{\bf 3}}]^{\vphantom{(}}_{\bf 3} (cu)_{\bar{\bf
3}}$.  The diquark $(cu)$ and antitriquark $\bar c (ud)$ achieve a
substantial separation before hadronization must occur, providing a
qualitative explanation for the suppression of the measured widths
compared to available phase space.  The $P_c^+$ states in this picture
form isospin doublets with neutral, as-yet undiscovered partners.
States with the observed $J^P$ quantum numbers can easily be
accommodated in this scheme, and suggest the potential for discovery
of numerous additional related states in the future.

\begin{acknowledgments}
I am deeply indebted to S.~Brodsky for his insights and encouragement,
and also thank M.~Karliner for interesting discussions.  This work was
supported by the National Science Foundation under Grant Nos.\
PHY-1068286 and PHY-1403891.
\end{acknowledgments}


\end{document}